\documentclass[12pt,epsf]{article}
\usepackage{graphicx}
\usepackage{wrapfig}
\begin{document}

\def\beq{\begin{eqnarray}}
\def\eeq{\end{eqnarray}}
\newcommand{\gsim}{ \mathop{}_{\textstyle \sim}^{\textstyle >} }
\newcommand{\lsim}{ \mathop{}_{\textstyle \sim}^{\textstyle <} }


\baselineskip 0.7cm

\begin{titlepage}

\begin{flushright}
IPMU 12-0056
\end{flushright}

\vskip 1.35cm
\begin{center}
{\large \bf
  $\theta_{13}$ in Neutrino Mass Matrix with the Minimal Texture
}
\vskip 1.2cm
Masataka Fukugita$^{a,b,c,}$\footnote{E-mail address: fukugita@icrr.u-tokyo.ac.jp}, 
Yusuke Shimizu$^{d,}$\footnote{E-mail address: shimizu@muse.sc.niigata-u.ac.jp}, \\
Morimitsu Tanimoto$^{d,}$\footnote{E-mail address: tanimoto@muse.sc.niigata-u.ac.jp}, 
 and  Tsutomu T. Yanagida$^{c}$, 
\vskip 0.4cm

$^a${\it \normalsize 
Institute for Advanced Study, Princeton, NJ08540, U. S. A. \\}
$^b${\it \normalsize 
Institute for Cosmic Ray Research, University of Tokyo, Kashiwa 277-8582, Japan \\}
$^c${\it \normalsize 
Kavli Institute for the Physics and Mathematics of the Universe, \\ University of Tokyo, Kashiwa 277-8583, Japan \\}
$^d${\it \normalsize 
Department of Physics, Niigata University, Niigata 950-2181, Japan \\}

\vskip 1.5cm

\abstract{We show that the neutrino mass matrix with the minimal
  texture, under the assumption that the neutrino is of the Majorana
  type, describes the neutrino mass and mixing that are empirically
  determined, including the finite $\theta_{13}$ that is measured
  recently at a high precision. This mass matrix only allows the
  normal hierarchy of the neutrino mass, excluding both inverse mass
  hierarchy and degenerate mass neutrinos. The current neutrino mixing
  data give the mass of the lightest neutrino to be between $0.7$ and
  $2.1$ meV, and predict the effective mass of double beta decay to
  lie in the range $m_{ee}=3.7-5.6$ meV.  The total mass of the three
  neutrinos should be $61\pm2$ meV.}

\end{center}
\end{titlepage}

\setcounter{page}{2}


Weinberg \cite{Weinberg} has pointed out, within the two generations of
quarks, that the relation is described between the quark mass and the
generation mixing if the quark mass matrix takes a simple
texture. Namely, with the 2$\times$2 mass matrix the vanishing (1,1)
element gives the mixing angle written as a square root of the quark
mass ratio of the down and the strange quarks, $(m_d/m_s)^{1/2}$ in
agreement with experiment. Fritzsch extended this argument to three
generations assuming the minimal texture of the 3$\times$3 matrix,
having off-diagonal (1,2) and (2,3) elements in addition to one
diagonal (3,3) matrix element retained \cite{Fritzsch}.

Following the discovery of the finite mass of neutrinos and, at the
same time surprisingly large mixing in the neutrino sector, the
present authors \cite{FTY93} proposed a mass matrix with the minimal
texture, where neutrinos are assumed to be of the Majorana type so
that the mixing angle is a quartic root of the neutrino mass ratio and
hence takes a large value unlike that in the quark sector. This matrix
was shown to give empirically derived mixing angles at a good
accuracy, and even more accurately as experiments develop \cite{FTY03}.
The novel prediction characteristics of this mass matrix is that the
mixing between the first and third components cannot be very small,
for which only an upper limit was known at the time. Using all
available data relevant to neutrino oscillation in the year 2003, the
$\theta_{13}$ is predicted to lie in the range sin$\theta_{13}\simeq
0.04-0.2$ ($\sin^22\theta_{13}=0.006-0.15$).  This means that
vanishingly small $\theta_{13}$, if measured to be so, would have falsified the
assumption of the minimal texture mass matrix.

$\theta_{13}$ was the last mixing angle to be measured to determine
the three Kobayashi-Maskawa angles in the neutrino sector. It was
within the last year that $\theta_{13}$ was measured to be finite,
$0.03<\sin^2(2\theta_{13})<0.28$ at the 90\% C.L.  (for normal
neutrino mass hierarchy) by the T2K experiment \cite{Abe11}. This 
was immediately confirmed by MINOS \cite{MINOS} and Double Chooz
\cite{DChooz} experiments.  Most recently it was measured at the Daya
Bay reactor neutrino experiment \cite{An12} at a high precision to
give ${\rm sin}^22\theta_{13} = 0.092\pm 0.016\pm 0.005$, which is 5.2
$\sigma$ to vanishing.

Motivated by the agreement of the minimal texture mass matrix with
experiments, we scrutinise further the model. We show that the
neutrino mass is well constrained within the present model, and there
are no uncertainties that allow the mass matrix to give the inverted mass
hierarchy or degenerate mass neutrinos.  This allows us to predict the
size of double beta decay without ambiguities.  It would be important
to see if this minimal mass matrix ansatz would be falsified in future
experiments.

Here, we briefly recapitulate our minimal texture hypothesis 
\cite{FTY93,FTY03} and the resulting consequences.
Our model consists of 
the mass matrices of the charged leptons and the Dirac neutrinos
of the form \cite{Fritzsch}
\begin{eqnarray}
m_E = \left( \matrix{0 & A_\ell & 0 \cr A_\ell & 0 & B_\ell \cr
                     0 & B_\ell & C_\ell\cr        } \right)\ ,\qquad\quad
m_{\nu D} = \left( \matrix{0 & A_\nu & 0 \cr A_\nu & 0 & B_\nu \cr
                     0 & B_\nu & C_\nu \cr        } \right )\  , 
\end{eqnarray}
where each entry is complex. We assume that neutrinos are of 
the Majorana type and 
take, for simplicity, the right-handed Majorana 
mass matrix to be the unit matrix, as
\begin{eqnarray}
M_R = M_0 {\bf 1},
\end{eqnarray}
where $M_0$ is much larger than the Dirac neutrino mass scale.  We may
consider a more general case where $M_R$ has three different
eigenvalues.  In such a case, however, the differences in the
eigenvalues can be absorbed into the wave functions of the
right-handed neutrinos, which leads to the violation of the symmetric
matrix structure (the minimal structure) of the neutrino mass.
Therefore, it amounts to the increase of the complexity of the
structure in the mass matrix.

We remark that our neutrino mass is stable against radiative
corrections.  For the heavy right-handed neutrino of mass
$O(10^{10}){\rm GeV}$ the Yukawa coupling for the Dirac neutrino
mass is smaller than $10^{-2.5}$.  A calculation with the
renormalisation group equation (e.g., \cite{RGE}) gives the radiative
correction to be $10^{-6}$ relative to the leading term, which is
negligible in our argument.

We obtain the three light neutrino masses, $m_i$ ($i=1,2,3$),  
as
\begin{equation}
m_i=\left (U_\nu ^Tm_{\nu D}^TM_R^{-1}m_{\nu D}U_\nu \right )_i.
\end{equation}
The lepton mixing matrix is given by
\begin{eqnarray}
 U = U_\ell^\dagger \ Q \ U_\nu
\end{eqnarray}
where the expressions of  $U_\ell$ and  $\ U_\nu$ 
 are given in \cite{FTY03}, and  $Q$ is a phase matrix written  
\begin{eqnarray}
Q= \left(\matrix{ 1 & 0 & 0 \cr 0 & e^{i \sigma} & 0\cr  0 & 0 & e^{i \tau}
\cr } \right)\ ,
\end{eqnarray}
which is a reflection of phases contained in the charged lepton mass
matrix and the Dirac mass matrix of neutrinos.  The Majorana phases
can be neglected in the right-handed neutrino mass matrix, because, in
the presence of mass hierarchy of the left-handed neutrinos, the
effect of phases in the right-handed mass matrix is suppressed by the
mass hierarchy, so that the inclusion of phases changes the analysis
only a little.

Since the charged lepton masses are known, the number of parameters
contained in our model is six: $m_{1D},\ m_{2D}, \ m_{3D}$, $\sigma$,
$\tau$ and $M_0$, which are to be determined by empirical neutrino
masses and mixing angles.  We note that this is the minimum texture of
the $3\times3$ neutrino mass matrix in the sense that reducing one
more matrix element (i.e., letting A, B or C to zero) leads to the
neutrino mixing that is in a gross disagreement with experiment. An
antisymmetric mass matrix with 3 finite elements also leads to a gross
disagreement with experiment, as one can readily see.

The 
 lepton mixing matrix elements are written approximately,
\begin{eqnarray}
U_{e1} &\simeq & \left (\frac{m_2}{m_2+m_1}\right )^{1/4}+\left
 (\frac{m_1}{m_2}\right )^{1/4}
 \left (\frac{m_e}{m_\mu }\right )^{1/2}e^{i\sigma } \nonumber \\
U_{e2} &\simeq& -\left({m_1 \over m_2}\right)^{1/4}+
\left({m_e\over m_\mu}\right)^{1/2}e^{i\sigma} \cr
U_{\mu 1} &\simeq& \left({m_1\over m_2}\right)^{1/4}e^{i\sigma}
-\left({m_e\over m_\mu}\right)^{1/2} \cr
U_{\mu 3} &\simeq&  \left({m_2\over m_3}\right)^{1/4}e^{i\sigma}
-\left({m_\mu\over m_\tau}\right)^{1/2} e^{i\tau}\cr 
U_{\tau 2} &\simeq&  -\left({m_2\over m_3}\right)^{1/4}e^{i\tau}
+\left({m_\mu\over m_\tau}\right)^{1/2} e^{i\sigma}\cr
U_{e3} &\simeq&  \left({m_e\over m_\mu}\right)^{1/2}U_{\mu 3}+ 
\left({m_2\over m_3}\right)^{1/2} \left({m_1\over m_3}\right)^{1/4}
   \cr
U_{\tau 1} &\simeq&  \left({m_1\over m_2}\right)^{1/4}U_{\tau 2}
\label{appromixing}
\end{eqnarray}
where the charged lepton mass is denoted as $m_e$, $m_\mu$ and
$m_\tau$, and terms with $(m_e/m_\tau)^{1/2}$ are neglected.
Approximate characteristics of the mixing angles can be seen from
these expressions.

For instance, we see from these equations the relations between
neutrino masses and mixing angles, 
\begin{equation} 
|U_{e2}|\approx
\left (\frac{m_1}{m_2}\right )^{1/4}, \quad |U_{\mu 3}|\approx \left
(\frac{m_2}{m_3}\right )^{1/4}, \quad |U_{e3}|\approx \left
(\frac{m_2}{m_3}\right )^{1/2}\left (\frac{m_1}{m_3}\right )^{1/4}.
\end{equation} 
The relation among the mixing angles is
\begin{equation} 
|U_{e3}|\approx |U_{\mu 3}|^2|U_{e2}U_{\mu
3}|=|U_{\mu 3}|^3|U_{e2}|.  
\end{equation} 
With $|U_{\mu 3}|\sim
1/\sqrt{2}$ and $|U_{e2}|\sim 1/\sqrt{3}$ we see that 
$|U_{e3}|\sim 1/(2\sqrt{6})\simeq 0.2$, which means that $|U_{e3}|$
cannot be too small.

We emphasize that only the normal neutrino mass hierarchy is allowed in
our model, which allows us to predict uniquely, up to errors, the
effective mass that appears in double beta decay.  We can see that
this matrix does not allow the inverted hierarchy, which is seen easily from
the expression of $|U_{\mu 3}|$ that should empirically be around
$1/\sqrt{2}$~\cite{FTY03}.  It can also be seen that this matrix does
not allow degenerate neutrinos, since they require $|U_{e2}|$ close to
$1/\sqrt{2}$, which is excluded by experiment for $\theta_{12}$
(see Figure 9
below).


We now present our results using the accurate expression of the
lepton mixing angles given in \cite{FTY03}.
We take  \cite{An12,Schwetz:2011qt}
\begin{eqnarray}
&&\Delta m_{\rm atm}^2=(2.24-2.65)\times 10^{-3}~{\rm eV}^2, \quad 
\Delta m_{\rm sol}^2=(7.29-7.92)\times 10^{-5}~{\rm eV}^2,  \nonumber \\
&&\sin ^2\theta _{23}=0.40-0.62, \quad
\sin ^2\theta _{12}=0.29-0.34, 
\end{eqnarray}
where $\theta_{ij}$ are the lepton mixing angles defined in the standard manner,
and we leave out for the moment $\theta_{13}$ for the input.
$\Delta m_{\rm atm}^2$ and $\Delta m_{\rm sol}^2$ 
represent mass difference squares relevant to
atmospheric neutrino ($\theta_{23}$) and solar neutrino ($\theta_{12}$)
experiments.
We take the error at the 90\% confidence level throughout our analysis.
We take the lowest neutrino mass $m_1$ as a free parameter; $m_2$ and $m_3$
are given by $\Delta m_{\rm atm}^2$ and $\Delta m_{\rm sol}^2$.
The phase parameters $\sigma$ and  $\tau$ are constrained reasonably well
to give empirical  
$\sin ^2\theta _{23}$ and $\sin ^2\theta _{12}$ .


\begin{figure}[h!]
\begin{minipage}[]{0.45\linewidth}
\includegraphics[width=7.5cm]{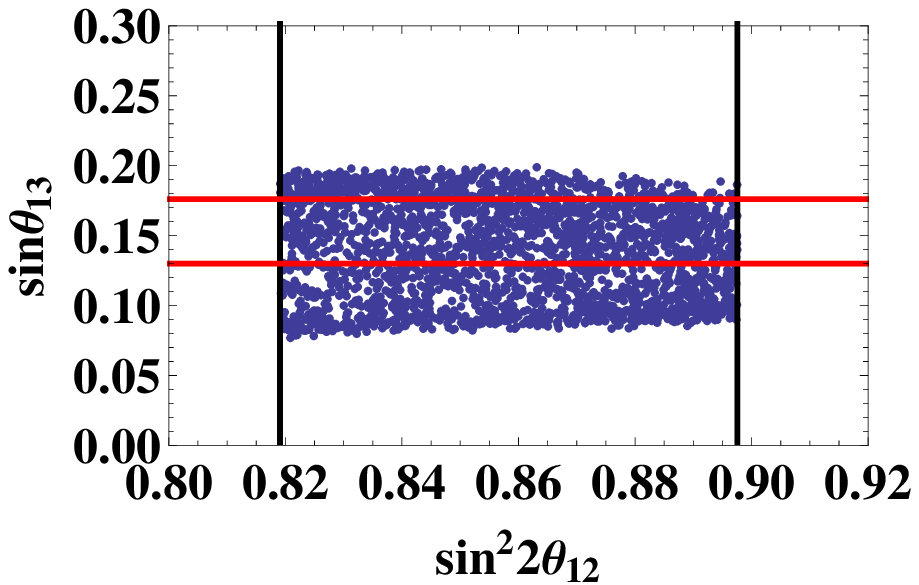}
\caption{Neutrino mixing $\sin ^22\theta _{13}$ as a function of
$\sin \theta _{13}$ plane, 
where the empirical value of $\sin \theta _{13}= |U_{e3}|$
is not used as the input.
The data points are marginalised
over the other free parameters of the model.
Horizontal and vertical lines denote 
90\% confidence limits for the two parameters.}
\end{minipage}
\hspace{5mm}
\begin{minipage}[]{0.45\linewidth}
\vspace{-2.5cm}
\includegraphics[width=7.5cm]{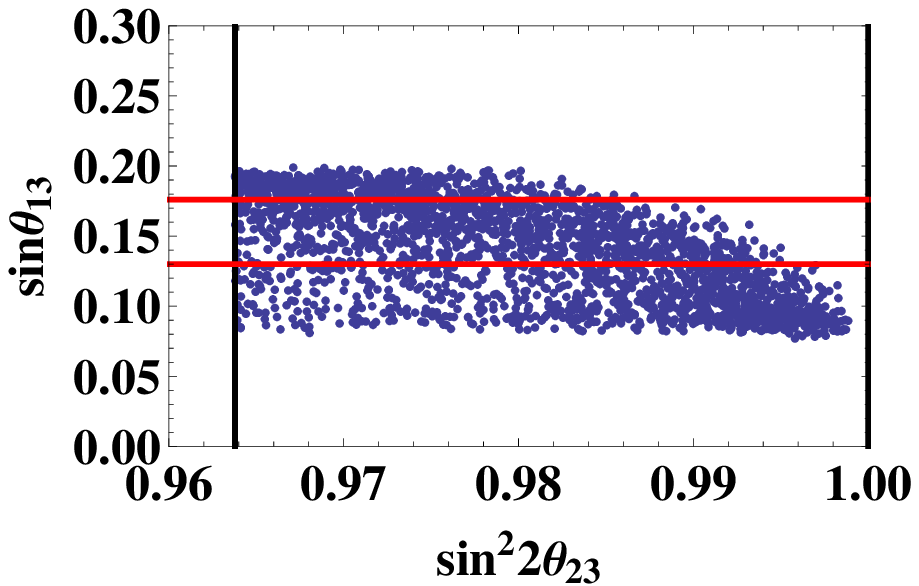}
\caption{Same as Figure 1 but  
shown as a function of $\sin ^22\theta _{23}$. }
\end{minipage}
\end{figure}

\begin{figure}[h!]
\begin{minipage}[]{0.45\linewidth}
\vspace{-1cm}
\includegraphics[width=7.5cm]{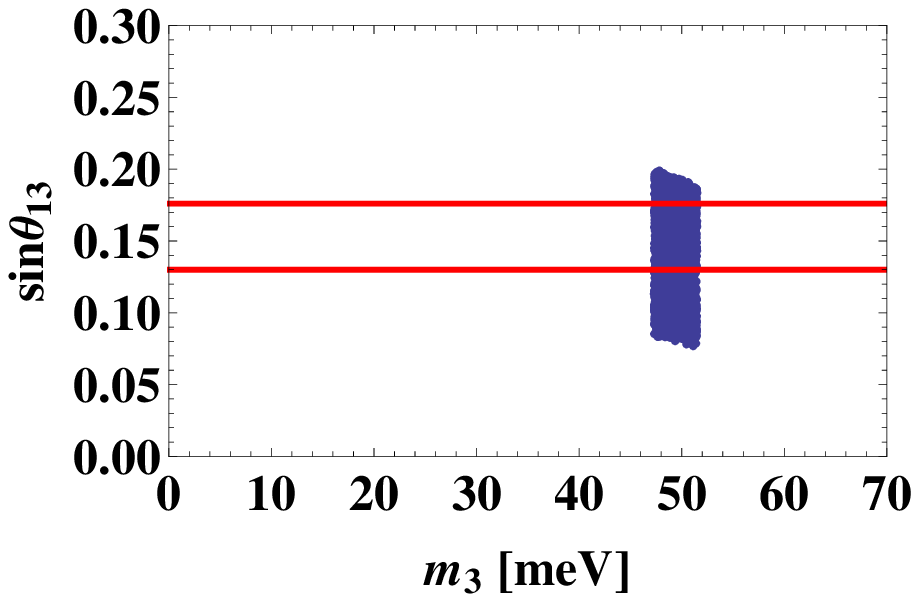}
\caption{$\sin \theta _{13}$ as a function of $m_3$.}
\end{minipage}
\hspace{5mm}
\begin{minipage}[]{0.45\linewidth}
\includegraphics[width=7.5cm]{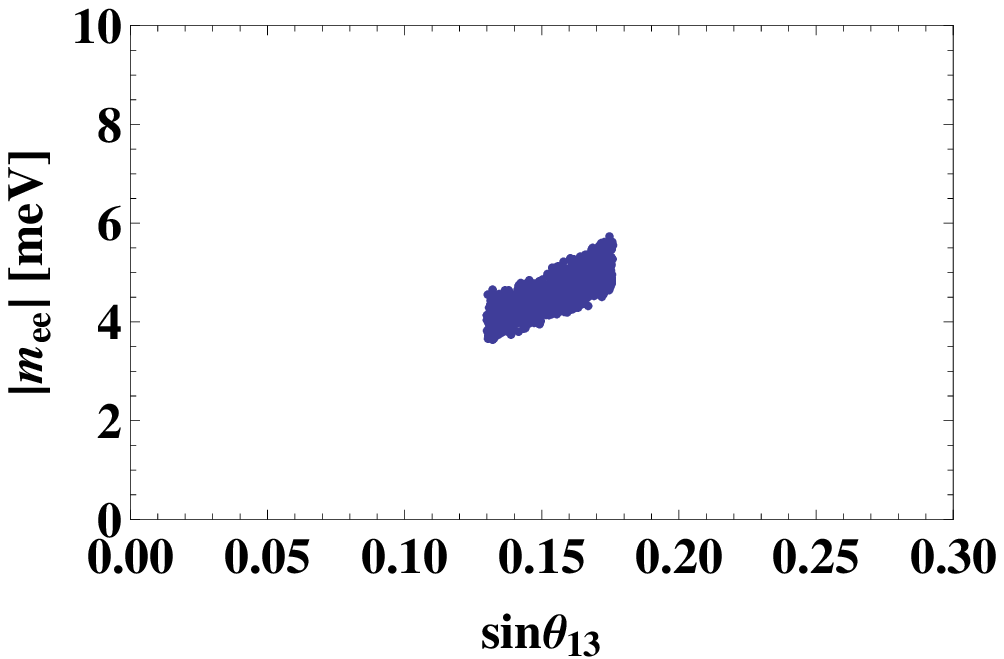}
\caption{Predicted effective mass $m_{ee}$ that appears in double beta
decay as a function of $\sin \theta _{13}$.}
\end{minipage}
\end{figure}

\begin{figure}[h!]
\begin{minipage}[]{0.45\linewidth}
\includegraphics[width=7.5cm]{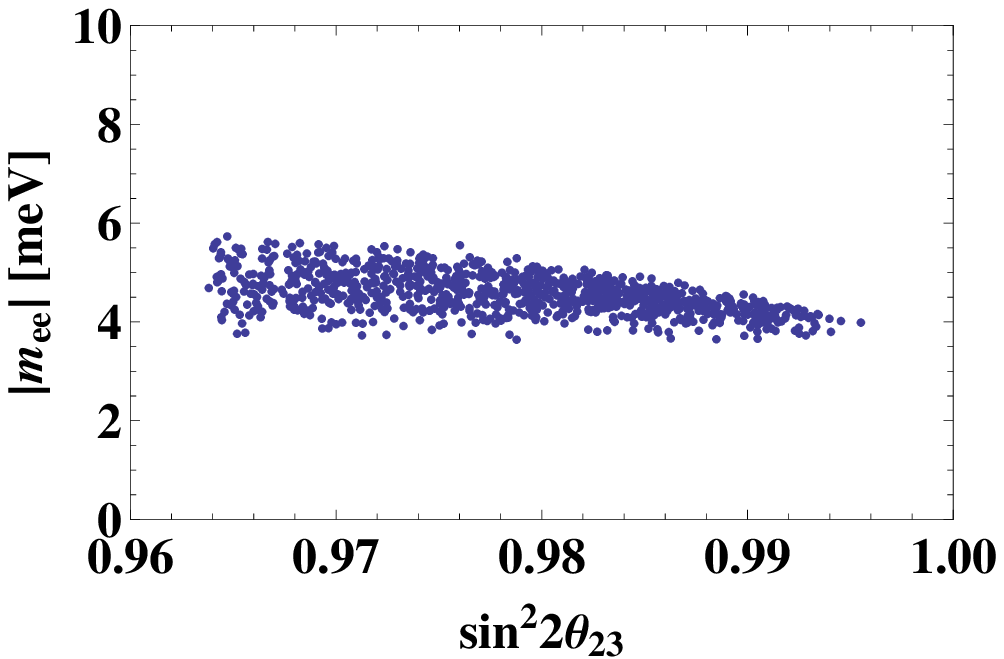}
\caption{ Effective mass $m_{ee}$ versus $\sin^2 \theta_{23}$.}
\end{minipage}
\hspace{5mm}
\begin{minipage}[]{0.45\linewidth}
\includegraphics[width=7.5cm]{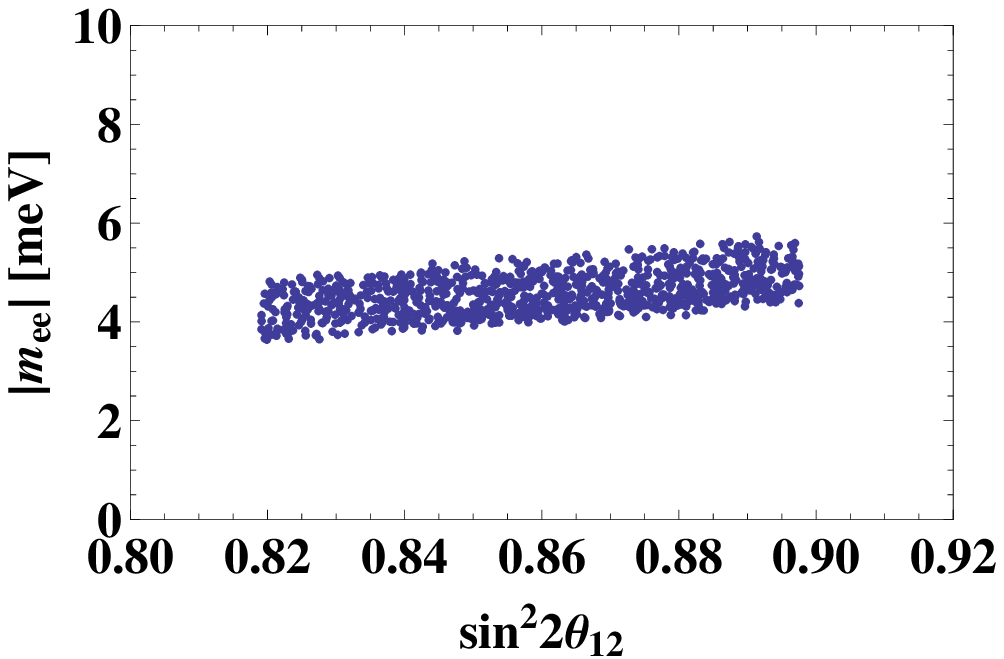}
\caption{Effective mass $m_{ee}$ versus $\sin^2 \theta_{12}$.}
\end{minipage}
\end{figure}

We consider the allowed region of   $|U_{e3}|=\sin \theta _{13}$.
Figures 1 and 2 show  $\sin \theta _{13}$ versus $\sin^22\theta _{12}$ 
and $\sin^22\theta _{23}$, respectively, where the other parameters
are marginalised.
The ranges of  $\sin^22\theta _{12}$ and $\sin^22\theta _{23}$ 
are cut at the boundary of the region 
experimentally allowed at the $90\%$ confidence level.
We stress that the predicted range of $\theta_{13}$ coincides with the
value obtained by the Daya Bay experiment.

We see that the maximum mixing, $\theta_{23}=\pi/4$  
(i.e., $\sin^2(2\theta _{23})=1)$, is excluded.
Figure 3 shows   $\sin\theta _{13}$ as a function of $m_3$.
It is noted that only the normal mass hierarchy, 
$m_3\gg m_2, \ m_1$, is allowed in our mass matrix.


Let us now impose
$\sin\theta_{13}=0.130-0.176$ (90\% C.L.)~\cite{An12} as an input parameter. 
The neutrino mass matrix is tightly constrained. 
The value of $m_1$ varies slightly with $\theta _{13}$ as seen in Figure 7 
and it must lie in the range $0.7-2.1$ meV. $m_2$ and $m_3$ are fixed 
by the oscillation experiments to lie in $m_2=8.6-9.1$ meV and $m_3=47-52$ meV,
respectively. 
Since the inverted mass hierarchy and the degenerate neutrinos are not allowed 
with our mass matrix, 
the effective mass that appears in neutrinoless double beta decay 
\begin{eqnarray}
m_{ee}=\sum _{i=1}^3m_iU_{ei}^2
\end{eqnarray}
is 
uniquely predicted:
\begin{equation}
|m_{ee}|=3.7-5.6 \ {\rm meV}.
\end{equation}
The dependence of $|m_{ee}|$ on the mixing angles is seen in 
Figures 4, 5 and 6.
We remark that the sum of the mass of three neutrinos, which is often of the
cosmological interest, is in the range,
\begin{equation}
\Sigma m_i=59-63~ {\rm meV}.
\end{equation}

\begin{wrapfigure}{r}{9.5cm}
\includegraphics[width=7.5cm]{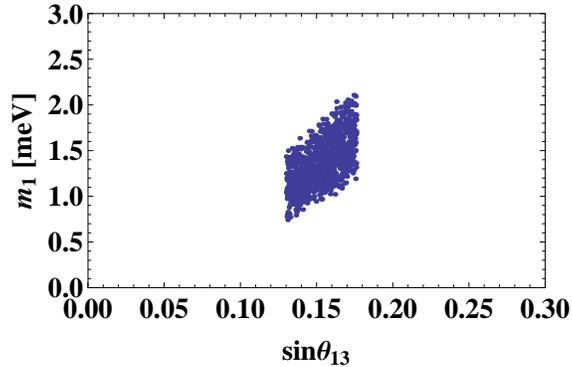}
\caption{The allowed value of $m_1$ as a function of $\sin \theta _{13}$.}
\end{wrapfigure}


We may also calculate the rephasing-invariant $CP$ violation measure $J_{CP}$  
defined by 
\begin{eqnarray}
J_{CP}={\rm Im}[U_{\mu 3}U_{\tau 3}^*U_{\mu 2}^*U_{\tau 2}].
\end{eqnarray}
in Figure 8 as a function of 
$\sin \theta _{13}$.
The figure shows that vanishing $J_{CP}$ is still marginally allowed, 
but some further improvement for the allowed range of $\theta_{13}$
will lead to non-vanishing $J_{CP}$. (If we would take
the one sigma range for $\theta_{13}$, vanishing $J_{CP}$ would not
be allowed any more.)

\begin{figure}[h!]
\begin{minipage}[]{0.45\linewidth}
\vspace{-2.1cm}
\includegraphics[width=7.5cm]{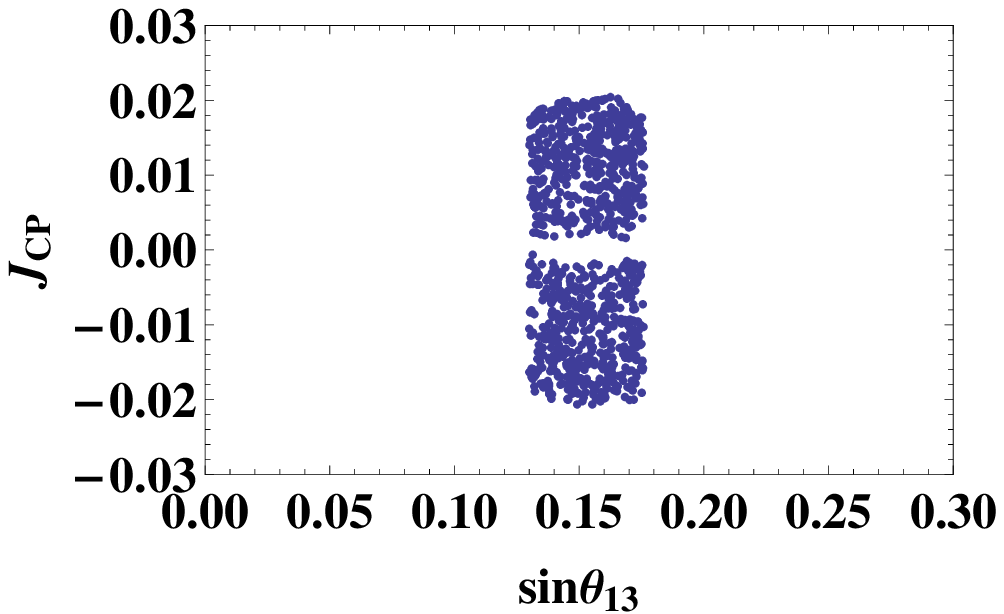}
\caption{rephasing-invariant CP violation measure 
$J_{\rm CP}$ as a function of $\sin \theta _{13}$.}
\end{minipage}
\hspace{5mm}
\begin{minipage}[]{0.45\linewidth}
\includegraphics[width=6.8cm]{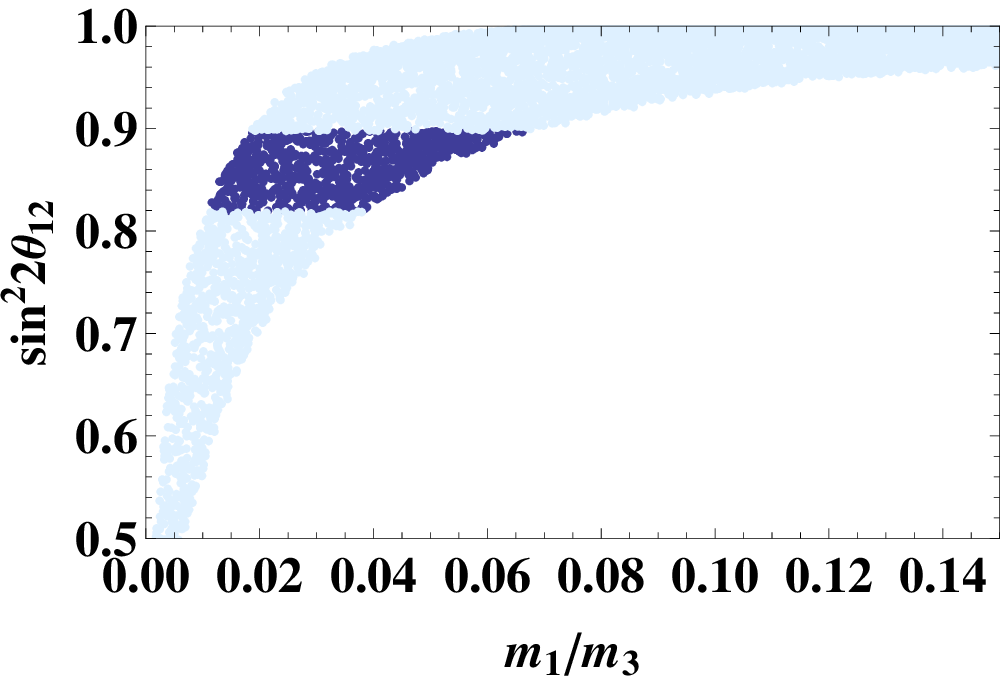}
\caption{$\sin^2\theta _{12}$ as a function of the mass ratio
  $m_1/m_3$. Data points are limited to the empirical 90\% confidence limit of
  $\theta_{12}$, and light shade indicates how the region would be
  extended if we do not limit the range of $\theta_{12}$.}
\end{minipage}
\end{figure}

We said earlier that the inverted hierarchy and the degenerate neutrino 
masses are 
not allowed. While the former is seen from the expression given above,
the latter is seen in Figure 9, where $\sin^2(2\theta_{12})$
is shown as a function of $m_1/m_3$. A large $m_1/m_3$ requires 
maximum mixing between the first and the second generation,
$\sin^2(2\theta_{12})= 1$, which is already not allowed by experiment.
This excludes the degenerate case.

We have shown that the neutrino mass and mixing derived from the
neutrino mass matrix of the minimal texture satisfy all 
experiments, without adding any further matrix elements or
extending the assumption. In particular, $\theta_{13}$ falls in the
middle of the range that is measured recently.  The matrix only allows
the normal hierarchy of the neutrino mass, excluding either inverse
hierarchy or degenerate mass cases.  This predicts the effective mass
of double beta decay to lie within the range $m_{ee}=3.7-5.6$ meV and
the total mass of the three neutrinos to be $61\pm2$ meV. 
Further improvement on the allowed range of $\theta_{13}$ would exclude
vanishing CP violation. It would be interesting to
see if this simple texture mass matrix model will be falsified in
future experiments when the accuracy is improved.

\vspace{1 cm}
\noindent
{\bf Acknowledgement}

M.F. is supported by the Monell Foundation in Princeton.  Y.S. and
M.T. are supported by Grants-in-Aid for Scientific Research of JSPS,
Nos.22.3014 and 21340055, respectively.



\begin{thebibliography}{99}

\bibitem{Weinberg} S. Weinberg, HUTP-77-A057, Trans.New York Acad.Sci.38:185-201, 1977.

\bibitem{Fritzsch}
 H. Fritzsch,  Phys. Lett. {\bf B73} (1978) 317; 
         Nucl. Phys. {\bf B115} (1979) 189.
\bibitem{FTY93}
  M.~Fukugita, M.~Tanimoto, T.~Yanagida,
  Prog.\ Theor.\ Phys.\  {\bf 89 } (1993)  263-268.
  
  
\bibitem{FTY03}
  M.~Fukugita, M.~Tanimoto, T.~Yanagida,
  Phys.\ Lett.\  {\bf B562 } (2003)  273-278.
  [hep-ph/0303177].

\bibitem{An12}
  F.~P.~An {\it et al.}  [DAYA-BAY Collaboration],
  arXiv:1203.1669 [hep-ex].

\bibitem{Abe11}
  K.~Abe {\it et al.}  [T2K Collaboration],
  Phys.\ Rev.\ Lett.\  {\bf 107} (2011) 041801
  [arXiv:1106.2822 [hep-ex]].



\bibitem{MINOS}
  P.~Adamson {\it et al.}  [MINOS Collaboration],
  Phys.\ Rev.\ Lett.\  {\bf 107} (2011) 181802
  [arXiv:1108.0015 [hep-ex]].

\bibitem{DChooz}
  Y.~Abe {\it et al.}  [DOUBLE-CHOOZ Collaboration],
  arXiv:1112.6353 [hep-ex].

\bibitem{RGE}
 C. Hagedorn, J. Kersten and M. Lindner, Phys. Lett. B{\bf 597} (2004) 63.


\bibitem{Schwetz:2011qt}
  T.~Schwetz, M.~Tortola, J.~W.~F.~Valle,
  New J.\ Phys.\  {\bf 13 } (2011)  063004.
  [arXiv:1103.0734 [hep-ph]].


\end{thebibliography}
\end{document}